\documentclass[aps,preprint,epsfig,rotate]{revtex4}
\usepackage{graphicx}
\usepackage{bm}
\usepackage{epsfig}
\input epsf
\begin{document}
%\begin{doublespace}
\title{On bound state computations in three- and four-electron atomic
       systems}

 \author{Alexei M. Frolov}
 \email[E--mail address: ]{afrolov@uwo.ca}

 \author{David M. Wardlaw}
 \email[E--mail address: ]{dwardlaw@uwo.ca}

\affiliation{Department of Chemistry\\
 University of Western Ontario, London, Ontario N6H 5B7, Canada}

\date{\today}

\begin{abstract}

A variational approach is developed for bound state calculations in three-
and four-electron atomic systems. This approach can be applied to determine,
in principle, an arbitrary bound state in three- and four-electron ions and
atoms. Our variational wave functions are constructed from four- and
five-body gaussoids which depend upon the six ($r_{12}, r_{13}, r_{14},
r_{23}, r_{24}, r_{34}$) and ten ($r_{12}, r_{13}, r_{14}, r_{15}, r_{23},
r_{24}, r_{25}, r_{34}, r_{35}$ and $r_{45}$) relative coordinates,
respectively. The approach allows one to operate with the different number
of electron spin functions. In particular, the trial wave functions for the
${}^1S$-states in four-electron atomic systems include the two independent
spin functions $\chi_1 = \alpha \beta \alpha \beta + \beta \alpha \beta
\alpha - \beta \alpha \alpha \beta - \alpha \beta \beta \alpha$ and $\chi_2
= 2 \alpha \alpha \beta \beta + 2 \beta \beta \alpha \alpha - \beta \alpha
\alpha \beta - \alpha \beta \beta \alpha - \beta \alpha \beta \alpha -
\alpha \beta \alpha \beta$. We also discuss the construction of variational
wave functions for the excited $2^3S$-states in four-electron atomic
systems.

PACS number(s): 31.10.-p, 31.15.ac and 31.15.ae

\end{abstract}
\maketitle
%\noindent \vspace{0.2in}

\section{Introduction}

In this study we consider the electronic structure of three- and
four-electron atoms and ions. In particular, we undertake variational
computations of the singlet ${}^1S$-states and triplet ${}^3S-$states in
various four-electron atoms and ions. Below, by an atomic system we mean a
system which contains a number of electrons and one heavy nucleus. In the
non-relativistic approximation used in this study the Hamiltonian of an
arbitrary $(A - 1)$-electron atomic system takes the form (see, e.g.,
\cite{LLQ})
\begin{equation}
 H = - \frac{1}{2} \Bigl[ \sum^{A-1}_{i=1} \nabla^2_i +
 \frac{1}{M} \cdot \nabla^2_{A} \Bigr] - \sum^{A-1}_{i=1} \frac{Q}{r_{i A}}
 + \sum^{A-2}_{i=1} \sum^{A-1}_{j=2(>i)} \frac{1}{r_{ij}} \label{e1}
\end{equation}
where $A$ is the total number of bodies in the atomic system. In a
three-electron atomic system one has $A = 4$ in Eq.(\ref{e1}). In this case
the subscripts 1, 2, 3 designate three electrons, while the subscript 4
denotes the positively charged atomic nucleus. For four-electron systems we
have $A = 5$ in Eq.(\ref{e1}) and the subscripts 1, 2, 3, 4 designate four
electrons, while the subscript 5 denotes the positively charged nucleus.
Note that the Hamiltonian, Eq.(\ref{e1}), and all equations that follow are
written in atomic units, where $\hbar = 1, m_e = 1, e = 1$. Also, in
Eq.(\ref{e1}) $\nabla_i = (\frac{\partial}{\partial x_i},
\frac{\partial}{\partial y_i}, \frac{\partial}{\partial z_i})$ is the
gradient operator of the $i-$th particle ($i = 1, 2, \ldots, A$). The
notation $r_{ij}$ stands for the $(ij)-$relative distance/coordinate between
$i-$th and $j-$th particles, i.e. $r_{ij} = \mid {\bf r}_i - {\bf r}_j \mid
= r_{ji}$, where ${\bf r}_i$ are the Cartesian coordinates of the $i-$th
particle. Also, in Eq.(\ref{e1}) the notation $M$ stands for the mass of the
central (heavy) nucleus, i.e. $M \gg 1$.

The main goal of this work is to discuss some important details of bound
state calculations of three- and four-electron atoms and ions. In
particular, special attention will be given to the correct symmetrization of
the four-electron trial wave functions which include more than one
independent (electron) spin function, something not considered elsewhere in
the modern literature. Formally, our main goal is to determine highly
accurate solutions of the Schr\"{o}dinger equation $H \Psi = E \Psi$ where
$H$ is the Hamiltonian, Eq.(\ref{e1}), while $E$ is the eigenvalue (= total
energy) of the considered bound state and $\Psi$ is the corresponding wave
function. It is clear that the permutation symmetry of the total wave
function $\Psi$ must be different for the singlet and triplet bound states
in four-electron system. The explicit construction of trial wave functions
with the correct permutation symmetry between all three or all four
electrons is the principal part of any accurate variational calculation of
such atomic systems.

In general, bound state computations of three-, four- and many-electron
atomic systems with the use of a number of different spin functions (which,
however, correspond to the same value of the total (electron) spin $S_e$ and
its $z-$projection $(S_e)_z$) are significantly more complicated than
analogous calculations for two-electron helium-like atoms and ions. On the
other hand, there is an obvious similarity in calculations of three- and
four-electron atomic systems. Indeed, for the ground (doublet) $1^2S$-states
in three-electron atoms and ions one needs to use the two different spin
functions. The same number of spin functions is needed for accurate
computations of the ${}^1S$-states in four-electron atomic systems.

Note that in any four-electron atom and/or ion all bound states are
separated into two series of states: singlet states with the total electron
spin $S_e$ equals zero, i.e. $S_e = 0$; and triplet states with the total
electron spin $S_e$ equals unity, $S_e = 1$. In all previous works only the
ground singlet $1^1S-$state was considered. Accurate computations of the
triplet states in four-electron atomic systems have been performed for a
very few ions/atoms \cite{FroWa1}, \cite{BeT}. Moreover, in almost all
previous computations of the singlet state in four-electron (Be-like) atoms
and ions only one spin function $\chi_1 = \alpha \beta \alpha \beta + \beta
\alpha \beta \alpha - \beta \alpha \alpha \beta - \alpha \beta \beta \alpha$
was used. Here and everywhere below in this study $\alpha$ and $\beta$ are
the spin-up and spin-down single electron functions, i.e. $\hat{\sigma}_z
\alpha = \frac12 \alpha$ and $\hat{\sigma}_z \beta = - \frac12 \beta$. The
second independent spin function $\chi_2 = 2 \alpha \alpha \beta \beta + 2
\beta \beta \alpha \alpha - \beta \alpha \alpha \beta - \alpha \beta \beta
\alpha - \beta \alpha \beta \alpha - \alpha \beta \alpha \beta$ has been
ignored in almost all modern accurate computations of the singlet states in
four-electron atomic systems. Bearing this in mind we wanted to develop the
method which can be used to perform bound state computations for the
${}^1S-$singlet and ${}^3S-$triplet bound states in arbitrary four-electron
atoms and ions. Our method is not restricted with respect to the number of
spin functions included. It works equally well in cases, when one, two,
three and even more independent spin functions are used.

\section{Variational wave functions}

A central feature of any variational method is the construction of trial
wave functions $\Psi$ with the correct permutation symmetry. In general,
such a trial wave function must include all electron and nuclear
coordinates. Accurate wave functions explicitly depend upon all scalar
interparticle coordinates $r_{ij} = \mid {\bf r}_i - {\bf r}_j \mid$ in the
atomic system. The use of a large number of relative coordinates complicates
the explicit symmetrization of the trial wave functions. Another
complication follows from the presence of different (independent) terms in
the spin part of the total wave function. For instance, let us discuss the
case of the singlet ${}^1S-$state in four-electron atomic systems. To
compute this state in this study we shall use the two following independent
spin functions written in the form $\chi_1 = \alpha \beta \alpha \beta +
\beta \alpha \beta \alpha - \beta \alpha \alpha \beta - \alpha \beta \beta
\alpha$ and $\chi_2 = 2 \alpha \alpha \beta \beta + 2 \beta \beta \alpha
\alpha - \beta \alpha \alpha \beta - \alpha \beta \beta \alpha - \beta
\alpha \beta \alpha - \alpha \beta \alpha \beta$. These two functions obey
the following relations
\begin{equation}
 {\bf S}^2 \chi_k = 0 \; \; \; , \; \; \; ({\bf S}_z) \chi_k = 0 \; \; \; ,
 \; \; \; \langle \chi_i \mid \chi_k \rangle = D_k \delta_{ik} \label{e2}
\end{equation}
where $k = 1, 2$; $D_k$ are the normalization factors of the spin functions
and ${\bf S} = {\bf s}_1 + {\bf s}_2 + {\bf s}_3 + {\bf s}_4$ is the total
electron spin of the four-electron system.

The total wave function of the ${}^1S(L = 0)$-state of the four-electron
beryllium-like atom/ion is written in the form (see, e.g., \cite{Fish},
\cite{FroWa2} and references therein)
\begin{equation}
 \Psi_{L=0} = \psi_{L=0}(A; \bigl\{ r_{ij} \bigr\}) \cdot \chi_1 +
 \phi_{L=0}(B; \bigl\{ r_{ij} \bigr\}) \cdot \chi_2 \label{psi2}
\end{equation}
where $\psi_{L=0}(A; \bigl\{ r_{ij} \bigr\})$ and $\phi_{L=0}(B; \bigl\{
r_{ij} \bigr\})$ are the two independent radial parts (= spatial parts) of
the total wave function. For the wave function, Eq.(\ref{psi2}), one finds
${\bf S}^2 \Psi = 0$ and $S_z \Psi = 0$, where ${\bf S} = {\bf s}_1 + {\bf
s}_2 + {\bf s}_3 + {\bf s}_4$ is the total electron spin of the
four-electron system. The radial parts $\psi_{L=0}(A; \bigl\{ r_{ij}
\bigr\})$ and $\phi_{L=0}(B; \bigl\{ r_{ij} \bigr\})$ of the total wave
function Eq.(\ref{psi2}) in this work are represented in the form \cite{KT}
\begin{eqnarray}
 \psi_{L=0}(A; \bigl\{ r_{ij} \bigr\}) = {\cal P}_1 \sum_{k=1}^{N_A} C_{k}
 \cdot \exp(- \sum_{ij} \alpha^{(k)}_{ij} r^2_{ij}) \label{gaus} \\
 \phi_{L=0}(B; \bigl\{ r_{ij} \bigr\}) = {\cal P}_2 \sum_{k=1}^{N_B} {\cal
 C}_{k} \cdot \exp(- \sum_{ij} \beta^{(k)}_{ij} r^2_{ij}) \label{gaus2}
\end{eqnarray}
where $N_A$ and $N_B$ are the number of basis functions used, $C_{k}$ and
${\cal C}_k$ are the linear parameters of the variational expansions,
Eq.(\ref{gaus}) and Eq.(\ref{gaus2}), while $\bigl\{ r_{ij} \bigr\}$ is
the set of relative coordinates which are needed for complete description of
five-body systems. The notations $\alpha^{(k)}_{ij}$ and $\beta^{(k)}_{ij}$
designate the non-linear parameters associated with the $r_{ij}$ relative
coordinate in the $k-$th basis function in the $\psi$ and $\phi$ radial
functions, respectively (see, Eqs.(\ref{gaus}) - (\ref{gaus2})). For all
beryllium-like ions and atoms considered in this study the notation
$\bigl\{ r_{ij} \bigr\}$ stands for ten relative coordinates $r_{12},
r_{13}, r_{14}, r_{15}, r_{23}, r_{24}, r_{25}, r_{34}, r_{35}$ and
$r_{45}$. The radial basis functions in Eqs.(\ref{gaus}) - (\ref{gaus2}) are
called the five-body gaussoids of the ten relative coordinates. This name
was used in \cite{KT} where these basis functions were invented for nuclear
few-body systems.

The main advantage of the radial function defined in Eqs.(\ref{gaus}) -
Eqs.(\ref{gaus2}) follows from the fact that the formulas for all matrix
elements do not depend explicitly upon the total number of particles in the
system. In other words, these formulas are essentially the same for three-,
four-, five- and many-body systems and are discussed in the next Section.

The notations $A$ and $B$ in Eq.(\ref{gaus}) and Eq.(\ref{gaus2}) mean that
there are two different sets of non-linear parameters. Note that each of
the basis functions in Eq.(\ref{gaus}) and Eq.(\ref{gaus2}) contains ten
non-linear parameters which are optimized independently. The summation over
$(ij) = (ji)$ in the exponents of Eq.(\ref{gaus}) and Eq.(\ref{gaus2}) is
taken over all possible different pairs of particles. In general, the radial
basis functions are not orthogonal to each other. The projectors ${\cal
P}_1$ and ${\cal P}_2$ produce the trial wave functions with the correct
permutation symmetry between all electrons (see below). The symbol $L$ in
Eq.(\ref{gaus}) is used for the total angular momentum of the considered
system. For the ground state of any Be-like system we always have $L$ = 0
and the total electron spin of such states equals zero. Furthermore, the
parity of these states in the four-electron systems is even. With respect to
this, these states are often designated as the ${}^1S^{e}$ states, or
$1^1S^{e}$ states.

For the triplet ${}^3S-$states in four-electron atomic systems one also
finds the two independent spin functions $\chi_1 = \alpha \beta \alpha
\alpha - \beta \alpha \alpha \alpha$ and $\chi_2 = 2 \alpha \alpha \beta
\alpha - \beta \alpha \alpha \alpha - \alpha \beta \alpha \alpha$. Note that
there are, in fact, the two independent triplets of the spin functions, i.e.
the total number of electron spin functions equals six (6 = 2 $\times$ 3).
The functions $\chi_1$ and $\chi_2$ mentioned above correspond to the values
$S_e = 1$ and $(S_e)_z = 1$. By using the explicit form of these two spin
functions one can construct the four remaining spin functions with $S_e = 1$
and $(S_e)_z = 0$ (two functions) and $S_e = 1$ and $(S_e)_z = -1$ (two
functions). In actual calculations of internal atomic structure we can
restrict ourselves to the use of the two spin functions $\chi_1$ and
$\chi_2$ only. All six spin functions are needed only in some special cases,
e.g., if an external magnetic field is present.

It is interesting to find that the total variational wave function $\Psi$ of
three-electron atoms and ions is also represented in the form of
Eq.(\ref{psi2}). In this case $\chi_1 = \alpha \beta \alpha - \beta \alpha
\alpha$ and $\chi_2 = 2 \alpha \alpha \beta - \beta \alpha \alpha - \alpha
\beta \alpha$, while the two independent radial parts depend upon six
relative coordinates $r_{12}, r_{13}, r_{23}, r_{14}, r_{24}, r_{34}$ (here
the indexes 1, 2, 3 mean three electrons, while 4 stands for the nucleus).
The radial basis functions can also be chosen in the form of
Eq.(\ref{gaus}) and Eq.(\ref{gaus2}), i.e. in the form of four-body
gaussoids. In this case each of the radial basis functions contains six
non-linear parameters. Note also that for each of the spin functions we have
$S^2 \chi_i = \frac34 \chi_i$ and $(S)_z \chi_i = +\frac12 \chi_i$ (here $i$
= 1, 2). The explicit constructions of the variational wave functions for
three- and four-electron atomic systems is discussed below.

\section{Matrix elements}

Actual computation of the matrix elements with the four- and five-body
gaussoids Eq.(\ref{gaus}) and Eq.(\ref{gaus2}), is based on analytical
formulas derived elsewhere (see, e.g., \cite{KT}, \cite{FroWa}). It was
mentioned already in \cite{KT} that the explicit expressions for all matrix
elements needed for the solution of the Schr\"{o}dinger equation depend upon
the total number of particles $A$ in the system as a numerical parameter. In
other words, the expressions for matrix elements in three-electron systems
coincide with the corresponding formulas for matrix elements obtained for
four- and five- and many-electron systems. Moreover, in some few-body
systems one of the electrons can be replaced by another particle, e.g., by
$\mu^-$ muon, but such a replacement does not change the explicit formulas
for matrix elements. Below, the symbol $A$ designates the total number of
particles (i.e. bodies) in the system. The $A-$particle atomic system
includes the $(A - 1)-$electron subsystem plus one heavy nucleus. The
$A-$particle muonic atom/ion contains the $(A - 2)-$electron subsystem, one
negatively charged muon $\mu^{-}$ and one heavy nucleus.

The explicit formulas for all matrix elements needed in computations of the
$A$-particle atomic systems can be presented in a very brief form with the
use of the following compact notation
\begin{equation}
 \langle \alpha \mid = \langle \alpha^{(k)} \mid = exp( -\sum_{i>j=1}^{A}
 \alpha^{k}_{ij} \cdot r^{2}_{ij})
 \; \; \; and \; \; \;
 \mid \beta \rangle = \mid \beta^{(\ell)} \rangle = exp( -\sum_{i>j=1}^{A}
 \beta^{\ell}_{ij} \cdot r^{2}_{ij}) \label{e6}
\end{equation}
where $A$ is the total number of particles in the system. In this notation
the symbols $\mid \beta \rangle$ and $\mid \alpha \rangle$ (or $\langle
\alpha \mid$ and $\langle \beta \mid$) stand for the radial basis functions.
These notations are different from the symbols $\alpha$ and $\beta$ used in
other Sections of this study to designate the spin-up and spin-down
functions.

In the notation defined in Eq.(\ref{e6}) the overlap matrix element $\langle
\alpha \mid \beta \rangle$ is written in the form
\begin{eqnarray}
 \langle \alpha \mid \beta \rangle = \langle \alpha^{(k)} \mid
 \beta^{(\ell)} \rangle = \pi^{\frac{3 \cdot (A-1)}{2}} \cdot
 D^{-\frac{3}{2}} \label{e7}
\end{eqnarray}
where $D$ is the determinant of the $(A-1) \times (A-1)-$matrix $\hat{B}$
with the matrix elements
\begin{eqnarray}
 b_{ii} = \sum_{j \neq i}^{A} (\alpha^{k}_{ij} + \beta^{\ell}_{ij}), \;
 \; i \neq j = 1, 2, \ldots , A - 1 \label{eq8} \\
 b_{ij} = -(\alpha^{k}_{ij} + \beta^{\ell}_{ij}), \; \; i \neq
 j = 1, 2, \ldots , A-1 \nonumber
\end{eqnarray}
In particular, the explicit expression for the ($k,\ell$) matrix element of
the overlap matrix $\hat{S}$ in the case of $A = 5$ is the $4 \times 4$
matrix $\hat{B}$ with matrix elements $b_{ij}$ defined in Eq.(\ref{eq8}).
Analytical and/or numerical computations of the determinant of this matrix
and all its first order derivatives is straightforward.

The formula for the appropriate matrix elements of the potential energy can
be written in the form:
\begin{eqnarray}
 \sum_{(ij)} \langle \alpha \mid V(r_{ij}) \mid \beta \rangle =
 \frac{4}{\sqrt{\pi}} \langle \alpha \mid \beta \rangle \sum_{ij}
 \int_{0}^{+\infty} V\Bigl( x \sqrt{\frac{D_{ij}}{D}}
 \Bigr) \cdot \exp(-x^{2}) \cdot x^{2} dx
\end{eqnarray}
where $D_{ij} = \frac{\partial D}{\partial \alpha_{ij}} = \frac{\partial
D}{\partial \beta_{ij}}$, while $(ij)$ = $(ji)$ = (12), (13), (23), (14),
(24), (34) for the three-electron atom/ion ($A$ = 4) and $(ij)$ = $(ji)$ =
(12), (13), (14), (15), $\ldots$, (35), (45) for the four-electron atom/ion
($A$ = 5). The explicit expressions for various interparticle potentials
often used in bound state calculations can be found in \cite{KT}. The
integral in the last formula is computed analytically in many actual cases,
including the case of Coulomb, Yukawa-type, exponential, oscillator, and
other potentials. The matrix elements of the kinetic energy take the form
(in atomic units)
\begin{equation}
 \langle \alpha \mid T \mid \beta \rangle = \frac{3}{2 D} \Bigl[
 \sum_{ijk=1}^{A} \frac{\alpha_{ik} \beta_{jk}}{m_{k}}
 (D_{ik} + D_{jk} - D_{ij}) \Bigr] \langle \alpha \mid \beta \rangle
 \label{eq10}
\end{equation}
where $m_{i} (i = 1, 2, \ldots, A)$ are the masses of the particles and $i
\neq j \neq k$ in Eq.(\ref{eq10}). The explicit formulas for matrix elements
of other operators written in the basis of many-dimensional gaussoids can be
found elsewhere (see, e.g., \cite{KT}, \cite{FroWa}).

If all formulas needed for matrix elements of the potential and kinetic
energies are known, then the solution of the incident Schr\"{o}dinger
equation is reduced to the following generalized eigenvalue problem
\begin{equation}
 \sum^{N}_{\beta=1} \Bigl( H_{\alpha,\beta} - E \cdot S_{\alpha,\beta}
 \Bigr) C_{\beta} = 0 \label{geneq}
\end{equation}
for $\alpha = 1, \ldots, N$, where $N$ is the total number of basis
functions used. Here $H_{\alpha,\beta} = T_{\alpha,\beta} +
V_{\alpha,\beta} = \langle \alpha \mid T \mid \beta \rangle + \langle
\alpha \mid V \mid \beta \rangle$ is the Hamiltonian matrix, while
$T_{\alpha,\beta} = \langle \alpha \mid T \mid \beta \rangle$ and
$V_{\alpha,\beta} = \langle \alpha \mid V \mid \beta \rangle$ are the
matrices of the kinetic and potential energies, respectively. The
$S_{\alpha,\beta} = \langle \alpha \mid \beta \rangle$ matrix in
Eq.(\ref{geneq}) is the overlap matrix, Eq.(\ref{e7}). For non-orthogonal
basis sets the overlap matrix is a typical dense matrix, i.e. all elements
of such a matrix differ, in general, from zero. Moreover, it can be shown
that the overlap matrix $\langle \alpha \mid \beta \rangle$ is a symmetric,
positively defined matrix. This means that all eigenvalues of the overlap
matrix are positive.

\section{Antisymmetrization of the trial wave functions}

Let us consider the antisymmetrization of the trial wave functions and
related antisymmetrization of the corresponding matrix elements derived in
the previous Section. As mentioned above the correct antisymmetrization is a
central part of the construction of explicitly correlated, trial wave
functions. In general, such a wave function depends upon all
electron-nuclear and electron-electron coordinates. In two-electron atoms
and ions the antisymmetrization of the total wave function is a trivial
problem, since the wave function of the two-electron system is always
represented as a product of a radial and two-electron spin functions.
Moreover, only singlet and triplet spin functions are possible in any
two-electron atom and/or ion. The singlet states have spin function $\chi_1
= \alpha \beta - \beta \alpha$, while triplet states have three spin
functions $\chi^{(1)}_2 = \alpha \alpha, \chi^{(2)}_2 = \alpha \beta + \beta
\alpha, \chi^{(3)}_2 = \beta \beta$. For the singlet spin function one finds
$S^2 \chi_1 = 0, S_z \chi_1 = 0$, while for the triplet spin functions we
have $S^2 \chi^{(i)}_2 = 1 (1 + 1) \chi^{(i)}_2 = 2 \chi^{(i)}_2$ and $S_z
\chi^{(i)}_2 = \kappa_i \chi^{(i)}_2$, where $\kappa_1$ = 1, $\kappa_2$ = 0
and $\kappa_3$ = -1, respectively. Here and below $S$ and $S_z$ are the
total electron spin of the system and its $z-$projection.

Note that the singlet $\chi_1$ spin function is antisymmetric upon the
electron variables. Therefore, its product with a symmetric radial function
produces a function which is completely antisymmetric upon all electron
variables. It is clear that such a function can be considered as a total
wave function with the correct permutation symmetry between two electrons.
For triplet states the corresponding radial function must be antisymmetric
upon all electron (spatial) coordinates.

In contrast with two-electron systems, the antisymmetrization of
three-electron wave functions is a significantly more complex process, since
for the same spin state one finds not one, but a number of different
independent spin functions. This statement is true even for the doublet
states with $S = \frac12$ (i.e. $2 S + 1= 2$) and $S_z = \pm \frac12$ in any
three-electron atomic system. In actual computations, such spin functions
are usually chosen to be orthogonal to each other. For instance, any
variational expansion written for the doublet ${}^2S$-states in a
three-electron atomic system must include the two independent spin functions
$\chi_1 = \alpha \beta \alpha - \beta \alpha \alpha$ and $\chi_2 = 2 \alpha
\alpha \beta - \beta \alpha \alpha - \alpha \beta \alpha$. The total wave
function for the ground doublet $1^2S(L = 0)-$state of the three-electron
atomic system is written in the form (see, e.g., \cite{Lars}, \cite{King})
\begin{eqnarray}
 \Psi_{L=0} = \psi_{L=0}(A; \bigl\{ r_{ij} \bigr\}) (\alpha \beta \alpha
 - \beta \alpha \alpha) + \phi_{L=0}(B; \bigl\{ r_{ij} \bigr\}) (2 \alpha
 \alpha \beta  - \beta \alpha \alpha - \alpha \beta \alpha) \label{psiX}
\end{eqnarray}
where $\psi_{L=0}(A; \bigl\{ r_{ij} \bigr\})$ and $\phi_{L=0}(B; \bigl\{
r_{ij} \bigr\})$ are the two independent spatial parts (= radial parts) of
the total wave function. The notations $A$ and $B$ mean that the two sets of
non-linear parameters associated with $\psi$ and $\phi$ are optimized
independently.

Note that for each of these two spin functions $\chi_1$ and $\chi_2$ the
two following equations are obeyed:
\begin{equation}
 {\bf S}^2 \chi_k = S (S + 1) \chi_k  = \frac34 \chi_k \; \; \; , \; \; \;
 ({\bf S})_z \chi_k = \frac12 \chi_k \label{Casim}
\end{equation}
where $k$ = 1, 2 and ${\bf S} = {\bf s}_1 + {\bf s}_2 + {\bf s}_3$ is the
total electron spin of the three-electron system, while ${\bf S}_z$ is its
$z-$projection. The conditions Eq.(\ref{Casim}) indicate clearly that the
two spin functions $\chi_1$ and $\chi_2$ are equally important in this
variational method. Therefore, we cannot neglect any of these spin functions
$a$ $priori$. This means that in any of our calculations for three-electron
atomic system we have to make appropriate use of the two different radial
functions and two spin configurations $\chi_1$ and $\chi_2$. The explicit
construction of the trial wave functions for three-electron atomic systems
with two independent spin functions is more complicated problem than in the
case of one spin function and its solution is based on the method of
projection operators is discussed below.

\subsection{Three-electron atomic systems. Doublet states.}

Suppose our trial wave function for a three-electron atomic system is
written in the form of Eq.(\ref{psiX}). In real applications, however, only
those trial functions are accepted which have the correct permutation
symmetry between all identical particles, i.e. electrons. This means that
the two terms in the right-hand side of Eq.(\ref{psiX}) must be completely
antisymmetric upon spin and spatial coordinates of the three electrons, i.e.
upon the indexes 1, 2 and 3 in our notation. In other words, we must have
$\hat{{\cal A}}_e \Psi = - \Psi$, where $\Psi$ is given by Eq.(\ref{psiX})
and $\hat{{\cal A}}_e$ is the three-particle (= electron) antisymmetrizer
\cite{PeTr}, \cite{Hein}
\begin{equation}
 {\hat{\cal A}}_e = \frac16 (\hat{e} - \hat{P}_{12} - \hat{P}_{13} -
 \hat{P}_{23} + \hat{P}_{123} + \hat{P}_{132}) \label{AS}
\end{equation}
Here $\hat{e}$ is the identity permutation, while $\hat{P}_{ij}$ is the
permutation of the $i$-th and $j$-th particles. Analogously, the operator
$\hat{P}_{ijk}$ is the permutation of the $i$-th, $j$-th and $k$-th
particles. The same notations are used everywhere below in the text.

By using the three-particle antisymmetrizer, Eq.(\ref{AS}), we can construct
a trial wave function of the correct permutation symmetry. In reality, we
need the matrix elements with the correct permutation symmetry, rather than
the wave function itself. Let us describe the approach which allows one to
obtain properly symmetrized matrix elements. First, note that the
expectation value of an arbitrary completely symmetric operator $W$ is
written in the form
\begin{equation}
 \langle {\hat{\cal A}}_e \sum \psi_i(A_i; \bigl\{ r_{ij} \bigr\}) \chi_i
 \mid W \mid {\hat{\cal A}}_e \sum \psi_j(A_j; \bigl\{ r_{ij} \bigr\})
 \chi_j \rangle \label{EV}
\end{equation}
where $\chi_i$ are the spin functions ($i = 1, \ldots, N_s$), where the
$\hat{{\cal A}}_e$ operator is defined in Eq.(\ref{AS}). The spin functions
$\chi_1$ and $\chi_2$ are assumed to be orthogonal to each other, i.e.
$\langle \chi_i \mid \chi_j \rangle = \delta_{ij}$. The notations
$\psi_i(A_i; \bigl\{ r_{ij} \bigr\})$ stand for the corresponding radial
functions which depend upon all relative coordinates $\bigl\{ r_{ij}
\bigr\}$ and non-linear parameters $A_i $. These radial functions can be
arbitrary, i.e. they are not necessarily orthogonal to each other. The
operator $W$ is a differential operator written in the relative coordinates.
It is assumed to be completely symmetric in respect to all inter-electron
permutations.

Now note that completely symmetric operator $W$ commutes with the
${\hat{\cal A}}_e$ operator, Eq.(\ref{AS}). Moreover, the operator
${\hat{\cal A}}_e$ is an orthogonal projector \cite{Halm}, i.e. $({\hat{\cal
A}}_e)^2 = {\hat{\cal A}}_e$ and $({\hat{\cal A}}_e)^* = {\hat{\cal A}}_e$,
where the symbol $B^*$ means the operator conjugate to the operator $B$. If
the operator $W$ does not depend upon spin variables, then by using these
properties of the ${\hat{\cal A}}_e$ operator one can reduce the formula,
Eq.(\ref{EV}), to the following form
\begin{equation}
 \sum_i \sum_j \langle \psi_i(A_i; \bigl\{ r_{ij} \bigr\}) \mid W \mid
 {\hat{\cal A}}_e \psi_j(A_j; \bigl\{ r_{ij} \bigr\}) \rangle
 \langle \chi_i \mid {\hat{\cal A}}_e \mid \chi_j \rangle \label{ME1}
\end{equation}
This expectation value can be re-written into another form with the use of
the following matrix notation
\begin{equation}
 [\Theta(W)]_{ij} = \langle \chi_i \mid {\hat{\cal A}}_e \mid \chi_j \rangle
 [W {\hat{\cal A}}_e]_{ij} \label{PE1}
\end{equation}
where $[ \ldots ]_{ij}$ designates the $(ij)-$matrix element of the
corresponding matrix. The matrix elements of this operator ($\Theta(W)$)
computed on any basis set of spatial three-electron wave functions have the
correct permutation symmetry between all identical particles (electrons).
The dimension of the $[ \Theta(W) ]_{ij}$ matrix equals to the number of
spin functions used in calculations.

First, let us compute the matrix elements of the ${\hat{\cal A}}_e$
operator, where ${\hat{\cal A}}_e$ designates the complete three-particle
antisymmetrizer, Eq.(\ref{AS}). Based on Eq.(\ref{PE1}) this operator
can be written in the form $\Theta(\hat{e})$, where $\hat{e}$ is the unit
operator. Note that operator ${\hat{\cal A}}_e$, Eq.(\ref{AS}), is written
in the form
\begin{equation}
 {\hat{\cal A}}_e = \sum_{abc} s_{abc} \hat{P}_{abc} \label{e18}
\end{equation}
where $s_{abc}$ is an integer number, while $\hat{P}_{abc}$ are the
interparticle permutations in the system of three identical particles (see
Eq.(\ref{AS})). The sum in Eq.(\ref{e18}) is computed over all interparticle
permutations possible in three-body systems, Eq.(\ref{AS}). The
$(ij)-$matrix element of the ${\hat{\cal A}}_e$ operator in our basis is
\begin{equation}
 [{\hat{\cal A}}_e]_{ij} = \sum_{abc} s_{abc} \langle \chi_i \mid
 \hat{P}_{abc} \mid \chi_j \rangle \hat{P}_{abc}
\end{equation}
In the case of the $(ij)-$matrix element of the $\Theta(W)$ operator one
finds
\begin{equation}
 [\Theta(W)]_{ij} = \sum_{abc} s_{abc} \langle \chi_i \mid \hat{P}_{abc}
 \mid \chi_j \rangle W \hat{P}_{abc} = W \Bigl[ \sum_{abc} s_{abc} \langle
 \chi_i \mid \hat{P}_{abc} \mid \chi_j \rangle \hat{P}_{abc} \Bigr]
 \label{PE2}
\end{equation}
Note that both the expectation value $\langle \chi_i \mid \hat{P}_{abc} \mid
\chi_j \rangle$ and $s_{abc}$ are the integers for all $abc$, while the
operator $P_{abc}$ is a projector which acts on the spatial coordinates of
three electrons ($a \rightarrow b \rightarrow c$). In other words, the
${\hat{\cal A}}_e$ operator is represented as the finite sum of all spatial
permutations $\hat{P}_{abc}$ with integer coefficients equal to the products
of $s_{abc}$, Eq.(\ref{e18}), and the $\langle \chi_i \mid \hat{P}_{abc}
\mid \chi_j \rangle$ expectation values. The computation of all expectation
values $\langle \chi_i \mid \hat{P}_{abc} \mid \chi_j \rangle = \langle
\chi_i \mid \hat{P}_{abc} \chi_j \rangle$ can be considered as the
integration over electron spin coordinates.

Based on Eq.(\ref{PE2}) we can introduce the following operator ${\cal P}$
\begin{equation}
 {\cal P} = {\cal D} \sum_{abc} s_{abc} \langle \chi_i \mid \hat{P}_{abc}
 \mid \chi_j \rangle \hat{P}_{abc} \label{PE3}
\end{equation}
where ${\cal D}$ is the normalization constant. The numerical value of
${\cal D}$ in Eq.(\ref{PE3})is determined by the idempotiency of the ${\cal
P}$ operator, i.e. ${\cal P}^2 = {\cal P}$. The explicit use of this
operator substantially simplifies all following formulas. For instance, the
$(\alpha,\beta)-$matrix element of any arbitrary completely symmetric
operator $W$ can be written in the form
\begin{equation}
 \langle {\cal P} \alpha \mid W \mid {\cal P} \beta \rangle =
 \langle \alpha \mid {\cal P} W {\cal P} \mid \beta \rangle =
 \langle \alpha \mid W {\cal P} \mid \beta \rangle =
 \langle \alpha \mid W \mid {\cal P} \beta \rangle \label{e22}
\end{equation}
where $\alpha$ and $\beta$ are non-symmetric basis functions. This matrix
element has the correct permutation symmetry between all electrons.
Moreover, all expectation values constructed with the use of these matrix
elements, Eq.(\ref{e22}), also have the correct permutation symmetry
between all electrons. This is the main advantage of constructing the
orthogonal spatial projector ${\cal P}$ in explicit form.

In actual computations of the doublet ${}^2S$-states in three-electron
atomic systems after the integration over electron spin coordinates, one
finds the four following spatial projectors
\begin{eqnarray}
 {\cal P}_{\psi\psi} = \frac{1}{2 \sqrt{3}} \Bigl( 2 \hat{e} + 2
 \hat{P}_{12} - \hat{P}_{13} - \hat{P}_{23} - \hat{P}_{123} - \hat{P}_{132}
 \Bigr) \\
 {\cal P}_{\psi\phi} = \frac12 \Bigl( \hat{P}_{13} - \hat{P}_{23} +
 \hat{P}_{123} - \hat{P}_{132} \Bigr) \\
 {\cal P}_{\phi\psi} = \frac12 \Bigl( \hat{P}_{13} - \hat{P}_{23} +
  \hat{P}_{123} - \hat{P}_{132} \Bigr) \\
 {\cal P}_{\phi\phi} = \frac{1}{2 \sqrt{3}} \Bigl( 2 \hat{e} - 2
  \hat{P}_{12} + \hat{P}_{13} + \hat{P}_{23} - \hat{P}_{123} -
  \hat{P}_{132} \Bigr)
\end{eqnarray}
Here the indexes $\psi$ and $\phi$ correspond to the notations for radial
functions used in Eq.(\ref{psiX}). Each of these projectors produces matrix
elements between the two radial basis functions from Eq.(\ref{psiX}) with
the correct permutation symmetry. Note that the two projectors ${\cal
P}_{\psi\phi}$ and ${\cal P}_{\phi\psi}$ written above coincide with each
other. It can also be shown that the three projectors ${\cal P}_{\psi\psi},
{\cal P}_{\psi\phi}$ and ${\cal P}_{\phi\phi}$ are orthogonal to each other.
In actual computations only the upper triangles of the Hamiltonian and
overlap matrices are used. Therefore, only the three projectors ${\cal
P}_{\psi\psi}, {\cal P}_{\psi\phi}$ and ${\cal P}_{\phi\phi}$ are important
in computations of the bound doublet ${}^2S-$states in all three-electron
atomic systems.

The approach described above allows one to construct the spatial parts of
the total variational wave functions with the correct permutation symmetry
between all identical particles in the three-electron atomic system. In our
previous work we have also found (see, e.g., \cite{FroWa1}, \cite{FroWa2})
that the same approach works perfectly for all four-, five- and
many-electron systems. Moreover, the symmetry of the electron spin
functions can also be different, e.g., for the singlet and triplet states in
four-electron systems. Below, the variational wave functions for the singlet
and triplet states in the four-electron atomic systems are explicitly
constructed. The explicit formulas for the spatial parts of trial wave
functions are derived with the use of the corresponding spatial projectors.

\subsection{Four-electron atomic systems. Singlet states.}

Numerical computations of the bound states in four-electron atomic systems
include the non-trivial step of antisymmetrization of all electronic
variables, i.e. variables 1, 2, 3 and 4 in the trial wave function $\Psi$.
The variational wave function $\Psi$ of any singlet ${}^1S(L = 0)-$state in
four-electron atomic systems is represented in the form of Eq.(\ref{psiY}),
i.e.
\begin{eqnarray}
 \Psi_{L=0} = \psi_{L=0}(A; \bigl\{ r_{ij} \bigr\}) (\alpha \beta \alpha
 \beta + \beta \alpha \beta \alpha - \beta \alpha \alpha \beta - \alpha
 \beta \beta \alpha) \label{psiY} \\
 + \phi_{L=0}(B; \bigl\{ r_{ij} \bigr\}) (2 \alpha \alpha \beta \beta
 + 2 \beta \beta \alpha \alpha - \beta \alpha \alpha \beta
 - \alpha \beta \beta \alpha - \beta \alpha \beta \alpha - \alpha \beta
  \alpha \beta) \nonumber
\end{eqnarray}
where $\psi_{L=0}(A; \bigl\{ r_{ij} \bigr\})$ and $\phi_{L=0}(B; \bigl\{
r_{ij} \bigr\})$ are the two independent spatial parts (= radial parts) of
the total four-electron wave function. The notations $A$ and $B$ mean that
the two sets of non-linear parameters associated with $\psi$ and $\phi$ are
optimized independently. Such a trial wave function must be antisymmetric
upon all electron variables (or variables 1, 2, 3 and 4 in our notations),
i.e. ${\hat{\cal A}}_e \Psi = - \Psi$, where
\begin{eqnarray}
 {\hat{\cal A}}_e = \hat{e} - \hat{P}_{12} - \hat{P}_{13} - \hat{P}_{23}
 - \hat{P}_{14} - \hat{P}_{24} - \hat{P}_{34} + \hat{P}_{123} +
 \hat{P}_{132} + \hat{P}_{124} + \hat{P}_{142} + \hat{P}_{134}
 \label{e29a} \\
 + \hat{P}_{143} + \hat{P}_{234} + \hat{P}_{243} - \hat{P}_{1234} -
 \hat{P}_{1243} - \hat{P}_{1324} - \hat{P}_{1342} - \hat{P}_{1423} -
 \hat{P}_{1432} + \hat{P}_{12} \cdot \hat{P}_{34} \nonumber \\
 + \hat{P}_{13} \cdot \hat{P}_{24} + \hat{P}_{14} \cdot \hat{P}_{23}
 \nonumber
\end{eqnarray}
is the complete four-particle antisymmetrizer. Here $\hat{e}$ is the
identity permutation, while $\hat{P}_{ij}$ is the permutation of the
particles $i$ and $j$. Analogously, the operators $\hat{P}_{ijk}$ and
$\hat{P}_{ijkl}$ are the permutations of the particles $i, j, k$ and $i, j,
k, l$, respectively.

By using the procedure described in the previous Section we can find the
explicit formulas for the corresponding spatial projectors. In fact, by
using the explicit form of the trial wave function $\Psi$ constructed for
singlet states in four-electron systems, Eq.(\ref{psiY}), and by integrating
over electron spin coordinates one finds the four following spatial
projectors:
\begin{eqnarray}
 {\cal P}_{\psi\psi} = \frac{1}{4 \sqrt{3}} \Bigl( 2 \hat{e} +
 2 \hat{P}_{12} - \hat{P}_{13} - \hat{P}_{23} - \hat{P}_{14} - \hat{P}_{24}
 + 2 \hat{P}_{34} + 2 \hat{P}_{12} \hat{P}_{34} + 2 \hat{P}_{13}
 \hat{P}_{24} + 2 \hat{P}_{14} \hat{P}_{23} \label{e26} \\
 - \hat{P}_{123} - \hat{P}_{132} - \hat{P}_{124} - \hat{P}_{142}
 - \hat{P}_{134} - \hat{P}_{143} - \hat{P}_{234} - \hat{P}_{243}
 - \hat{P}_{1234} - \hat{P}_{1243} \nonumber \\
 + 2 \hat{P}_{1324} - \hat{P}_{1342} - \hat{P}_{1432} + 2
 \hat{P}_{1423} \Bigr) \nonumber \\
%%%%
 {\cal P}_{\psi\phi} = \frac{1}{4} \Bigl( \hat{P}_{13} -
  \hat{P}_{23} - \hat{P}_{14} + \hat{P}_{24} + \hat{P}_{123} -
  \hat{P}_{132} - \hat{P}_{124} + \hat{P}_{142} + \hat{P}_{134} -
  \hat{P}_{143} - \hat{P}_{234} + \hat{P}_{243} \\
 + \hat{P}_{1234} - \hat{P}_{1243} - \hat{P}_{1342} +
 \hat{P}_{1432} \Bigr) \nonumber \\
%%%%%
 {\cal P}_{\phi\psi} = {\cal P}_{\psi\phi} \\
%%%%%
 {\cal P}_{\phi\phi} = \frac{1}{4 \sqrt{3}} \Bigl( 2 \hat{e} -
 2 \hat{P}_{12} + \hat{P}_{13} + \hat{P}_{23} + \hat{P}_{14} + \hat{P}_{24}
 - 2 \hat{P}_{34} + 2 \hat{P}_{12} \hat{P}_{34} + 2 \hat{P}_{13}
 \hat{P}_{24} + 2 \hat{P}_{14} \hat{P}_{23} \label{e29} \\
 - \hat{P}_{123} - \hat{P}_{132} - \hat{P}_{124} - \hat{P}_{142}
 - \hat{P}_{134} - \hat{P}_{143} - \hat{P}_{234} - \hat{P}_{243}
 + \hat{P}_{1234} + \hat{P}_{1243} \nonumber \\
 - 2 \hat{P}_{1324} + \hat{P}_{1342} + \hat{P}_{1432} - 2
 \hat{P}_{1423} \Bigr) \nonumber
\end{eqnarray}
In reality, since ${\cal P}_{\psi\phi} = {\cal P}_{\phi\psi}$, one needs to
use only three such operators ${\cal P}_{\psi\psi}, {\cal P}_{\psi\phi}$ and
${\cal P}_{\phi\phi}$. The use of these three projectors for matrix elements
allows one to produce the matrix elements which have the correct permutation
structure between all four identical particles (electrons). Note that all
such matrix elements are computed only between the corresponding spatial
basis functions and do not include any spin function. The explicit formulas
for the complete set of singlet spatial projectors for four-electron atomic
systems, Eqs.(\ref{e26}) - (\ref{e29}), have not been presented in previous
publications. First bound state computations of four-electron atomic systems
with the use of completely correlated wave functions were performed by Sims
and Hagstr\"{o}m \cite{Sims}. Since then many authors have conducted such
calculations for singlet states in various four-electron systems (see, e.g.,
\cite{FroWa2} and references therein).

\subsection{Four-electron atomic systems. Triplet states.}

The trial wave function of the triplet ${}^3S$-state in the four-electron
atomic system can also be represented in the form with the two independent
spin functions $\chi_1 = \alpha \beta \alpha \alpha - \beta \alpha \alpha
\alpha$ and $\chi_2 = 2 \alpha \alpha \beta \alpha - \beta \alpha \alpha
\alpha - \alpha \beta \alpha \alpha$. The variational expansion takes the
form
\begin{equation}
 \Psi = \psi_{L=0}(A;\bigl\{ r_{ij} \bigr\}) \cdot (\alpha \beta \alpha
 \alpha - \beta \alpha \alpha \alpha) + \phi_{L=0}(B;\bigl\{ r_{ij} \bigr\})
 \cdot (2 \alpha \alpha \beta \alpha - \beta \alpha \alpha \alpha - \alpha
 \beta \alpha \alpha) \label{psiT2}
\end{equation}
where $\psi_{L=0}(A; \bigl\{ r_{ij} \bigr\})$ and $\phi_{L=0}(B;\bigl\{
r_{ij} \bigr\})$ are the radial parts (also called the spatial part) of the
total wave function. Here the notation $\bigl\{ r_{ij} \bigr\}$ designates
the complete set of fifteen interparticle (spatial) coordinates, while the
symbol $A$ and $B$ mean the corresponding sets of non-linear parameters.
Optimization of non-linear parameters in the $A-$ and/or $B$-sets is
performed independently from each other. The trial wave function,
Eq.(\ref{psiT}), contains two electron spin functions $\chi_1$ and $\chi_2$
which correspond to the $S = 1$ and $S_z = 1$ values, where $S$ and $S_z$
are the eigenvalues of the total electron spin and its $z-$projection, i.e.
${\bf S}^2 \chi_i = S (S + 1) \chi_i$ and ${\bf S}_z \chi_i = S_z \chi_i$.

For triplet states by using the explicit form of the $\chi_1$ and $\chi_2$
functions one can easily find the four other spin functions which correspond
to the $S = 1$ and $S_z = 0$ and $S = 1$ and $S_z = -1$ values. For
instance, in the case of $\chi_1 = \chi^{(+1)}_1$ spin function the two spin
functions $\chi^{(0)}_1 = \alpha \beta \alpha \beta + \alpha \beta \beta
\alpha - \beta \alpha \alpha \beta - \beta \alpha \beta \alpha$ and
$\chi^{(-1)}_1 = \alpha \beta \beta \beta - \beta \alpha \beta \beta$
correspond to the $S = 1, S_z = 0$ and $S = 1, S_z = -1$ values,
respectively.  The three spin functions $\chi^{(+1)}_1, \chi^{(0)}_1,
\chi^{(-1)}_1$ form a regular triplet of spin functions. An analogous
triplet of spin functions can be constructed for the $\chi_2$ spin function.
To describe experimental situations with no external magnetic field present
we need to use the spin functions associated with one value of $S_z$, say,
$S_z = 1$. Thus in this work we shall always choose $\chi_1 = (\alpha \beta
- \beta \alpha) \alpha \alpha = \alpha \beta \alpha \alpha - \beta \alpha
\alpha \alpha$ and $\chi_2 = 2 \alpha \alpha \beta \alpha - \beta \alpha
\alpha \alpha - \alpha \beta \alpha \alpha$ (see Eq.(\ref{psiT2}) above).

In calculations with such a trial wave function, Eq.(\ref{psiT2}), one needs
to know the explicit formulas for all three radial (or spatial) projectors
${\cal P}_{\psi\psi}$, ${\cal P}_{\phi\psi} (= {\cal P}_{\psi\phi})$ and
${\cal P}_{\phi\phi}$. However, numerical calculations using a wave function
with two spin functions are computationally intensive and were not attempted
in the current study. Accordingly, we did not attempt to derive the
associated projectors. Instead, we performed some computations of the
triplet states in four-electron atomic systems with the use of one spin
function $\chi_1 = \chi^{(+1)}_1 = \alpha \beta \alpha \alpha - \beta \alpha
\alpha \alpha$ only. The variational expansion Eq.(\ref{psiT2}) is now
written in the form
\begin{equation}
 \Psi = \psi_{L=0}(A;\bigl\{ r_{ij} \bigr\}) \cdot (\alpha \beta - \beta
 \alpha) \alpha \alpha = \psi_{L=0}(A;\bigl\{ r_{ij} \bigr\}) \cdot (\alpha
 \beta \alpha \alpha - \beta \alpha \alpha \alpha) \label{psiT}
\end{equation}
Now, we need to obtain the spatial part of the total wave function with the
correct permutation symmetry between all identical particles 1, 2, 3 and 4
(electrons). The corresponding spatial projector is obtained in this work
by calculating the explicit expression for the following spin expectation
value
\begin{eqnarray}
 {\cal P}_{\psi\psi} = {\cal C} \sum_{abcd} s_{abcd} \langle (\alpha \beta
 - \beta \alpha) \alpha \alpha \mid {\hat{P}}_{abcd} \mid (\alpha \beta
 - \beta \alpha) \alpha \alpha \rangle {\hat{P}}_{abcd}
\end{eqnarray}
where ${\cal C}$ is the normalization factor, while the integers $s_{abcd}$
are defined from the explicit form of the complete four-particle (or
four-electron) antisymmetrizer ${\hat{\cal A}}_e$ given in Eq.(\ref{e29a}).
After some algebra one finds the explicit formula for the corresponding
spatial projector
\begin{eqnarray}
 {\cal P}_{\psi\psi} = \frac{1}{2 \sqrt{6}} \Bigl( 2 \hat{e} +
 2 \hat{P}_{12} - \hat{P}_{13} - \hat{P}_{23} - \hat{P}_{14} - \hat{P}_{24}
 - 2 \hat{P}_{34} - 2 \hat{P}_{12} \hat{P}_{34} - \hat{P}_{123}
 - \hat{P}_{124} \label{e33} \\
 - \hat{P}_{132} - \hat{P}_{142} + \hat{P}_{134} + \hat{P}_{234}
 + \hat{P}_{243} + \hat{P}_{143} + \hat{P}_{1432} + \hat{P}_{1234}
 + \hat{P}_{1243} + \hat{P}_{1342} \Bigr) \nonumber
\end{eqnarray}
This projector creates the spatial part of an arbitrary matrix element
needed in bound state computations of the triplet ${}^3S-$states in an
arbitrary four-electron atomic system. Such a matrix element has the correct
permutation symmetry among all four identical particles (electrons).
Explicit formulas for the spatial projectors which correspond to the triplet
states have not been published previously.

By using the formulas presented above one can perform accurate computations
of the triplet bound states in various four-electron atomic systems. As
follows from our results of such calculations (see, e.g., \cite{FroWa1}) the
method described above allows one to determine various expectation values in
four-electron atomic systems to relatively high
numerical accuracy. In particular, such expectation values can be computed
for all positive and negative powers of relative coordinates $r_{ij}$. In
general, the expectation value of any regular function of the ten relative
coordinates $r_{ij}$ can be computed to very good numerical accuracy.
Analogous expectation values which contain delta-functions of the relative
coordinates and their products with the regular functions of the relative
coordinates also do not present any problem for numerical computations. Real
problems arise in computations of expectation values which include products
of delta-functions with the corresponding electron spin functions, e.g.,
$\langle \alpha_i \delta_{Ni} \rangle$ and $\langle \alpha_i \beta_j
\delta_{Ni} \delta_{Nj} \rangle$, where $N$ designates the nucleus, while
$i$ means $i$-th electron. The first expectation value $\langle \alpha_i
\delta_{Ni} \rangle$ represents the single-electron density of
$\alpha$-electrons on the atomic nucleus. Analogous expectation values can
be computed in the case of $\beta-$electrons. A very poor convergence of
similar expectation values means that another spin function must be
included in computations. In Eq.(\ref{psiT2}) such a spin function is called
the second (electron) spin function $\chi_2$.

\section{Generalization to the five- and six-electron atomic systems.}

The method described above allows one to construct the properly
antisymmetrized trial functions for three- and four-electron atomic systems.
Formally, our method can be generalized to the five, six and many-electron
atomic systems. However, its direct generalization is very difficult, since
the proper antysymmetrization of the trial basis functions and their linear
combinations becomes extremely difficult in the case of many-electron atoms
with $A - 1 \ge 5$ electrons, i.e. for $A \ge 6$. Here and everywhere below,
$A$ is the total number of particles in the system, while $A - 1$ is the
total number of electrons. This means to construct the trial wave functions
for five-, six- and many-electron atoms one needs to apply $A! (\ge 120)$
different interparticle permutations to the non-symmetrized basis function.
The presence of a very large number of terms in each wave function
drastically complicates the explicit expressions for the spatial projectors
mentioned above. For instance, such a spatial projector constructed for the
B-atom (five-electron atom) must include 120 different terms. Some of these
terms can be equal zero identically, but in any case the total number of
remaining terms is still very large. Therefore, it is important to develop
some effective methods which can be used to operate with a very large number
of terms in the trial wave functions. Our current hopes rely on the two
following methods. The first approach is based on the use of various
symbolic-algebra computational platforms such as Maple \cite{Mapl}. In this
approach all integrations over spin variables can be performed analytically.
The expressions for all spatial projectors are never written explicitly, but
they are used internally by this computational platform. Note also that for
some basis sets the action of any interparticle permutation
$\hat{P}_{abc\ldots}$ on the basis wave functions is reduced to the
permutation of the corresponding non-linear parameters in these functions.
In particular, this is the case for variational expansion defined by
Eqs.(\ref{gaus}) and (\ref{gaus2}). This means that actual permutations
of the non-linear parameters in the basis wave functions, Eqs.(\ref{gaus})
and (\ref{gaus2}), can always be applied instead of the permutation of the
relative coordinates. This drastically simplifies the explicit construction
of the completely symmetrized trial wave functions. The permutation of the
non-linear parameters in the basis wave functions can be combined with the
analytical integration over spin variables in the total wave function. This
can be used in the future methods.

The second approach is based on the relations which exist for the spin
functions in three-, four- and many-electron systems. For instance, the
second spin function $\chi_2$ used in numerical computations of the triplet
states of four-electron systems (see above) is obtained from the $\chi_2$
spin function known for the doublet states in three-electron systems.
Formally, we can write $\chi_2(1,2,3,4) = \chi_2(1,2,3) \alpha(4)$, where
$\alpha(4)$ is the spin function of the additional (= fourth) electron,
while symbols (1,2,3,4) and (1,2,3) designate the systems of four and three
electrons, respectively. The notation $\chi_2(1,2,3)$ stands for the second
spin function of the doublet state in three-electron system, while
$\chi_2(1,2,3,4)$ means the second triplet spin function of the
four-electron system. Analogous relation exists between another (first)
triplet spin function of the four-electron atomic systems,
$\chi_1(1,2,3,4)$, and $\chi_1(1,2,3)$ used above (see, e.g.,
Eq.(\ref{psiX})) for the doublet states in three-electron atoms/ions. By
studying this and other similar relations between spin functions we can
find some useful connections between the spatial projectors constructed for
three- and four-electron systems. This approach can also simplify methods
and algorithms which must be developed in the future for systems 5 or more
electrons.

\section{Numerical results}

To illustrate our method in applications to actual three- and four-electron
atomic systems let us briefly describe the results of variational
computations of bound states in the three-electron Be$^{+}$ ion in its
$1^2S$-state and four-electron Be atom in its $1^1S-$ and $2^3S$-states. For
simplicity all nuclear masses were assumed to be infinite in such
calculations. A separate group of calculations have been performed for the
$2^3S-$electron state in the six-body oxygen-muonic ion O$^{8+} \mu^{-}
e^{-}_4$, where $\mu^-$ is the negatively charged muon. This positively
charged ion ($q$ = + 3) is a well bound atomic system which contains the
composite `nucleus' (O$^{8+} + \mu^{-}$) with overall `nuclear' charge +7
and four atomic electrons. Below, we consider the ${}^{16}$O nucleus only.
In our calculations of the O$^{8+} \mu^{-} e^{-}_4$ ion we used $M$ =
29156.9457 $m_e$ for the nuclear mass of the oxygen-16 nucleus and $m_{\mu}$
= 206.768262 $m_e$ \cite{COD}, \cite{CRC}.

Numerical results of our computations can be found in Table I where one
finds the total energies $E$ and some other bound state properties expressed
in atomic units, where $\hbar$ = 1, $m_e$ = 1 and $e$ = 1. The electron
state of each atomic system is shown in the following brackets. In the case
of the O$^{8+} \mu^{-} e^{-}_4$ ion the notation $2^3S_{e}$ stands for the
$2^3S_{e}$-triplet electron state in this system. The muonic quasi-nucleus
O$^{8+} + \mu^{-}$ is in its ground $1^1S-$state. This is always assumed,
but not shown in our notation. Note also that for bound state properties
presented in Table I the index $e$ means the electron, while the index $N$
denotes the atomic nucleus. For each energy shown in Table I only 9 decimal
digits are presented. In general, optimization of the non-linear parameters
in variational expansion Eqs.(\ref{gaus}) and (\ref{gaus2}) always decreases
the total energies. On the other hand, small variations in a few last
decimal digits are not critically important for our present purposes.

As follows from Table I our method provides very good numerical accuracy for
doublet states in three-electron atoms and ions. On the other hand, this
method also works perfectly for singlet and triplet four-electron atomic
systems. It is very likely that the analogous procedure can be developed for
five-, six- and many-electron atomic systems. However, for atomic systems
with five and more electrons one finds a number of additional problems and
direct generalization of our method is very difficult (see discussion in the
previous Section).

Note also that the overall convergence rates of the radial variational
expansion, Eqs.(\ref{gaus}) - (\ref{gaus2}), for three- and four-electron
atomic systems are comparable with each other. It seems to be very strange,
but we need to remember that the number of non-linear parameters in each
basis function rapidly increases as the total number of bodies $A$ in
system increases. For four-electron atomic systems each of these basis
function contains 10 non-linear parameters (each of them varied
independently), while for three-electron atomic systems one finds only 6 such
parameters in each basis function. Briefly, this means that the overall
`flexibility' of the four-electron trial function is comparable with the
analogous `flexibility' of the three-electron trial functions.

\section{Conclusion}

We have considered the problem of accurate computations of bound states in
three- and four-electron atomic systems. The method developed in this study
allows one to construct variational wave functions for an arbitrary bound
state in three- and four-electron atomic systems. All such trial wave
functions have the correct permutation symmetry (with respect to all
permutations of identical particles, i.e. electrons). It is important to
note that in our method the total number of independent spin functions can
be varied. Numerical computations can start with the use of one electron
spin function only. The second spin function can be introduced later to
improve the overall convergence of the results. Our procedure can be
generalized to bound states with the non-zero angular momentum $L$. Such a
generalization is straightforward, but it requires extensive use of
additional notations, application of special methods developed in theory of
angular momentum and substantial explanations. Variational calculations of
the bound $P(L = 1)-$states in five-electron atomic systems will be
considered in our next study. The ground state of the B-atom is the $P(L =
1)-$state.

Our method is based on explicit constructions of the total wave functions
for various bound states in three- and four-electron atomic systems. The
unified procedure has been applied to an example of each type of system.
The central part of the procedure is the construction of spatial projectors
with the correct permutation symmetry between all identical particles
(electrons). This method was originally developed for three-electron atomic
systems by Larsson \cite{Lars} (see also \cite{King}). We generalized this
procedure to the cases when a number of different spin functions are used in
computations. In addition, we have constructed spatial projectors needed in
calculations of the singlet and triplet bound states in four-electron atomic
systems. Currently, the total energies and other properties of bound states
in four-electron atomic systems can be determined to the accuracy which is
better than the accuracy of old bound state calculations performed for
two-electron atomic systems \cite{BS}. In general, by using various
optimization strategies for non-linear parameters from the trial wave
functions one can obtain very accurate variational energies and highly
accurate wave functions. Such wave functions can be used in the following
accurate computations of different bound state properties, including various
relativistic and QED corrections.

\newpage
%
% TABLE I
%
   \begin{table}
    \caption{The non-relativistic energies and other properties determined
         for the $S-$states in some three- and four-electron atoms, ions
         and muonic ions (in atomic units).}
      \begin{center}
      \begin{tabular}{ccccc}
      \hline\hline
   & Be$^+$ ($1^2S$) & Be ($1^1S$) & Be ($2^3S$) &
     O$^{8+} \mu^{-} e^{-}_4$ ($2^3S_{e}$) \\
      \hline
  $E$                     & -14.3247627 & -14.6673323 & -14.4300595 & -6619.33457 \\
       \hline
 $\langle r^{-1}_{eN} \rangle$ & 2.65796 & 2.10684  & 2.03603  & 4.77655 \\

 $\langle r^{-1}_{ee} \rangle$ & 1.08200 & 0.72912  & 0.61933  & 1.26072 \\
       \hline
 $\langle r_{eN} \rangle$      & 1.03379 & 1.49297  & 2.63085  & 0.98798 \\

 $\langle r_{ee} \rangle$      & 1.75565 & 2.54516  & 4.70847  & 2.43483 \\
  \hline\hline
  \end{tabular}
  \end{center}
  \end{table}
\end{document}